\documentclass[pdflatex,sn-mathphys]{sn-jnl}

\usepackage{caption}
\usepackage{subcaption}
\usepackage{amsmath}
\usepackage{booktabs}

\usepackage{cleveref}
\usepackage{color}

\jyear{2023}%

\begin{document}
\title[MFML for Excited States]{Multi-Fidelity Machine Learning for Excited State Energies of Molecules}

\author[1]{\fnm{Vivin} \sur{Vinod}} 

\author[2]{\fnm{Sayan} \sur{Maity}} 

\author*[1]{\fnm{Peter} \sur{Zaspel}}\email{pzaspel@constructor.university}

\author*[2]{\fnm{Ulrich} \sur{Kleinekath\"ofer}}\email{ukleinekathoefer@constructor.university}

\affil[1]{\orgdiv{School of Computer Science and Engineering}, \orgname{Constructor University}, \orgaddress{\street{Campus Ring 1}, \postcode{28759} \city{Bremen}, \country{Germany}}}

\affil[2]{\orgdiv{School of Science}, \orgname{Constructor University}, \orgaddress{\street{Campus Ring 1}, \postcode{28759} \city{Bremen}, \country{Germany}}}

\abstract{
The accurate but fast calculation of molecular excited states is still a very challenging topic. For many applications, detailed knowledge of the energy funnel in larger molecular aggregates is of key importance requiring highly accurate excited state energies. To this end, machine learning techniques can be an extremely useful tool though the cost of generating highly accurate training datasets still remains a severe challenge. To overcome this hurdle, this work proposes the use of multi-fidelity machine learning where very little training data from high accuracies is combined with cheaper and less accurate data to achieve the accuracy of the costlier level. 
In the present study, the approach is employed to predict the first excited state energies for three molecules of increasing size, namely, benzene, naphthalene, and anthracene. The energies are trained and tested for conformations stemming from classical molecular dynamics simulations and from real-time density functional tight-binding calculations. It can be shown that the multi-fidelity machine learning model can achieve the same accuracy as a machine learning model built only on high cost training data while having a much lower computational effort to generate the data. The numerical gain observed in these benchmark test calculations was over a factor of 30 but certainly can be much higher for high accuracy data.
}


\maketitle

\section{Introduction}\label{Intro}
\begin{figure}[tb!]
    \centering
    \includegraphics[width=\textwidth]{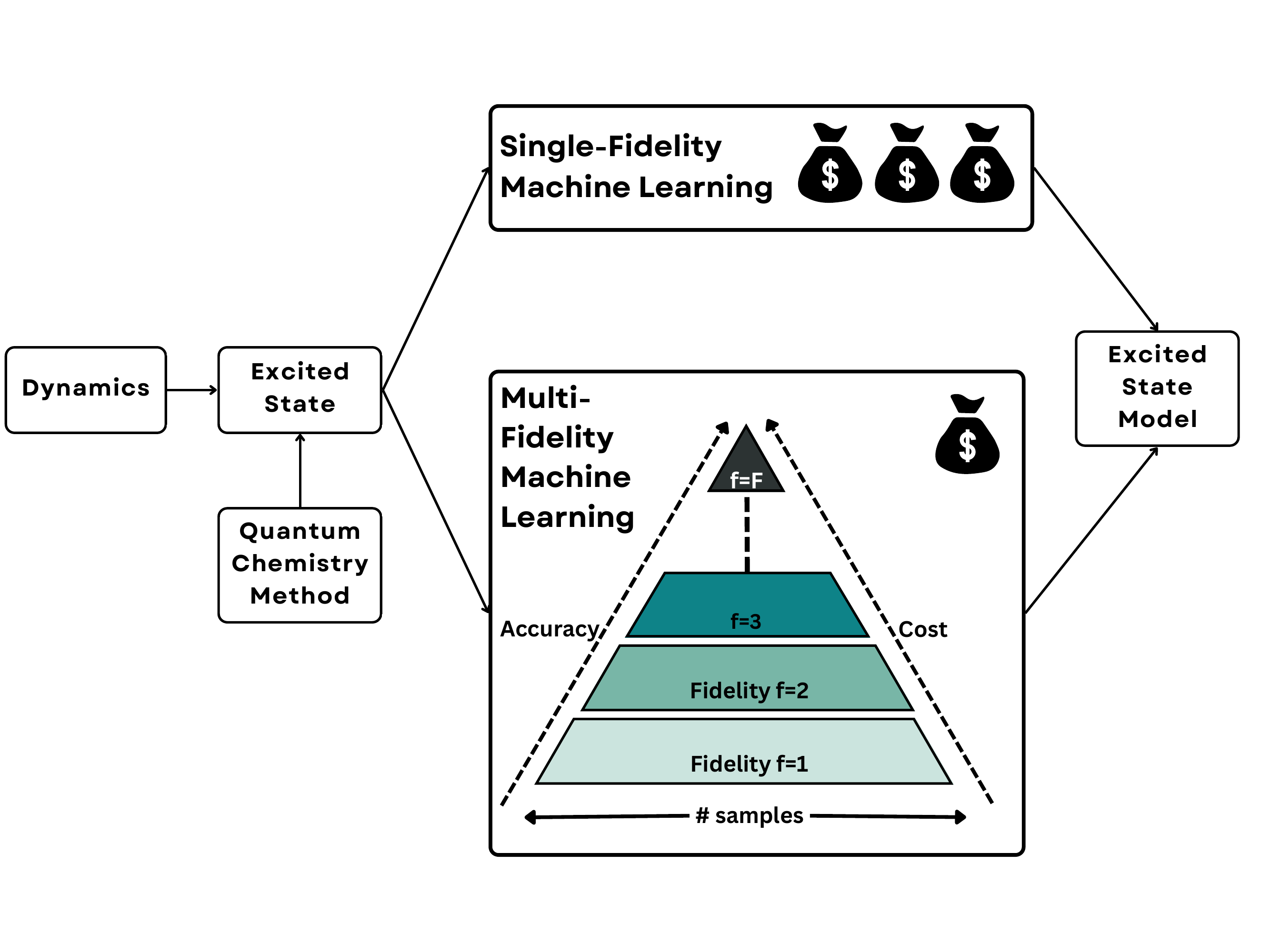}
    \caption{Multi-fidelity machine learning (MFML) based on kernel ridge regression significantly reduces the cost of training a machine learning model for the prediction of quantum chemistry properties, here, excited state energies. In contrast to the conventional single-fidelity machine learning method, the discussed method uses data from multiple fidelities with a few highly accurate (and costly) data samples and a growing number of less accurate (hence usually cheaper) data samples, thereby reducing the overall computational cost for the generation of the training data. This procedure helps to expedite the machine learning pipeline for predicting (first) excited state energies.}
    \label{fig_hierarchy}
\end{figure}

Excited states form the basis for understanding various photo-induced processes in physics, chemistry, and the life sciences. A detailed knowledge of their energies and properties is key in uncovering the secrets of the intricate working of many systems. Moreover, in many technical applications such as photovoltaics or light-emitting diodes, excited states play a key role as well. As one particular example, we would like to mention the collection of solar energy by light-harvesting complexes not only in plants and algae but also in some bacteria \cite{scho11a}. To be more specific, a recent model of a light-harvesting organelle, a chromatophore, includes more than 2000 pigment molecules. In order to determine the flow of excitation energy in such a system or to study its spectroscopic properties, one usually needs to determine the time evolution of the excited states for this large number of molecules \cite{zueh19a,cign22a,mait23a}. At the same time, however, the determination of the excited state energies needs to be rather accurate since small differences in energy can influence the direction of energy flow which might be crucial for a proper functioning of the biological process. Hence, for an accurate description of such systems, each single-point calculation usually comes with a high computational cost, which is amplified by the large number of those calculations. 

In recent years, machine-learning techniques have been applied to this area of excited state calculations, resulting in predictive models that are faster than conventional computational methods \cite{haes16a,chen20a,gupt21a,chen22a,cign23a,west20a,dral21a}. 
The computational effort to calculate molecular properties is shifted from an on-line calculation during the quantum chemistry computational run to an off-line phase, in which only the training data is generated, and machine learning models are trained. In such models, it has been commonly observed that the larger the number of training samples, the better the accuracy of prediction \cite{west20a,dral21a}. 
Since excited state calculations at high accuracy are expensive to perform, the cost of generating the training data imposes a demanding obstacle to train accurate machine learning models. 
Therefore, methods to reduce the necessity for numerous highly accurate but costly calculations to generate the necessary training data are needed.

The primary motivation in this work is thus to reduce the cost involved in generating the training data, without compromising on the accuracy of prediction. The number of training samples and the time to calculate individual training samples jointly contribute to this total cost. Currently, various ways exist to reduce the total cost of generating training data. These usually fall under the category of selecting optimal molecular conformations for training. For instance, active learning approaches shift the training data generation back to the on-line phase, where training samples are adaptively added to the training set based on estimators of the prediction error or the variance of the constructed model \cite{behler2015constructing, lin2020automatically}. In contrast, sampling techniques like the “\emph{de novo} exploration” of a potential energy surface \cite{bernstein2019novo} or an \emph{ab initio} random structure searching \cite{deringer2018data} select well distributed molecular samples in the off-line phase.
The $\Delta$-ML \cite{rama15a, dral22a} approach is another off-line phase method and adds a second training set for the same molecular conformations with either the same or a different chemical property, which is typically cheaper to compute. By only learning the difference between the cheaper and more expensive property, the approach results in a prediction error comparable to that of conventional machine learning methods but for a smaller training set size. 
It should be noted that the method still requires the same number of samples to be computed for the numerically cheaper and the more costly properties.
The $\Delta$-ML method has been used for the prediction of various quantum chemical properties such as potential energy surfaces \cite{nand21a}, band gaps \cite{zhan22a}, and excited state energies \cite{verm22a}.

In the multi-fidelity machine learning (MFML) method, also termed Combination Technique Quantum Machine Learning (CQML) \cite{zasp19a}, it is possible to combine sub-models that utilize a few training samples of the highest fidelity while using more samples from the cheaper fidelities to achieve the accuracy of a certain target fidelity. 
The MFML approach is a systematic generalization of the $\Delta$-ML method and exploits the correlations across multiple levels or fidelities in order to determine a certain property. In contrast to the two-level $\Delta$-ML approach, the MFML approach discussed in the present study uses several and not just two levels of calculations in order to enhance the gain in numerical efficiency. Moreover, the numerical efficiency is enhanced by decreasing the number of required training samples at higher levels of accuracy. Previously, the here discussed MFML method has been used for the prediction of atomization energies \cite{zasp19a}. 
Moreover, in Refs.~\cite{patra2020multi, Pilania2017} the related two-level multi-fidelity co-kriging approach is applied for the prediction of band gaps. 
In contrast, the hML formalism \cite{dral2020hierarchical} was used for high-accuracy PES reconstructions using multiple $\Delta$-ML models. The number of training samples for each of these models is optimized by a semi-automatic procedure and has been reported to reduce the numerical cost of the training set generation by a factor of 100 over a final model built with 8 different $\Delta$-ML models \cite{dral2020hierarchical}. In this respect, the hML approach can be seen as a specific case of the MFML method by fixing the training samples using a semi-automatic optimization. Apart from MFML being built with decreasing sizes of training samples at the higher fidelities, data on the lowest fidelity level is also replaced by an ML model, which eliminates the need to re-calculate the properties on this level during predictions.

The main aim of this work is to further develop and evaluate the MFML method for excited state energy calculations. Below, it will be shown that the MFML approach allows to drastically reduce the cost of the training data generation, while achieving the same prediction errors as classical machine learning models in the field. Thus, in the off-line phase, the number of costlier calculations is substantially reduced, as the numerical results in this work show. MFML combines sub-models that utilize a few training samples of the highest accuracy or \textit{fidelity} with sub-models using more samples from cheaper fidelities to achieve the accuracy of a certain target fidelity. The sub-models are all built using classical machine learning models. Although various models exist for machine learning in quantum chemistry, kernel ridge regression (KRR) and neural networks (NNs) are the two most predominantly used methods \cite{west20a}. The choice of the machine learning method for this research is KRR. 
One point is that KRR models are often considered easier to optimize for the prediction \cite{therrien2019impact, Westermayr2020NNKRR}. 
Furthermore, KRR is considered to be less prone to overfitting in comparison to NNs, where external steps like early-stopping and k-fold cross-validation are implemented to prevent overfitting \cite{west20a}. 
The regularly reported drawback of the KRR approach lies in the cubic scaling of solving a system of linear equations, but is less threatening to the present application since the desired low error is already achieved for a maximum kernel matrix size of $2^{13}=8192$.  

The use of machine learning for the prediction of quantum chemistry properties such as the first excited state energy requires that the Cartesian geometry of the molecules be transformed into some machine-learnable features. This transformation is achieved by \textit{representations} or \textit{molecular descriptors}.
The descriptors which encode the chemical and physical properties of the molecule \cite{RupCM, rupp2014machine, rupp2015machine} become the input to the machine learning model and are then used by the model to find a map between the descriptor and the property to be predicted. 
For this work, unsorted Coulomb Matrices (CMs) are used. This choice is due to their simplicity and robustness for the type of data used in this work. In the present case, the CMs are not row-sorted since (i) this work does not require invariance under permutations for the implementation as the models are built for individual types of molecules and not across the chemical space, (ii) the ordering of atoms is identical across individual frames of the trajectories for all molecules studied, and (iii) row-sorting the CMs is known to introduce discontinuities that are undesirable \cite{RupCM, krae20a}.

While the long-term interest is, for example, on light-harvesting complexes containing chlorophyll molecules in complex environments, it is pertinent to first establish the present method for smaller molecules in gas phase. This work benchmarks MFML and its effectiveness on three molecules of growing size, namely benzene, naphthalene, and anthracene.
The training data is based on time-dependent trajectories calculated by classical molecular dynamics (MD) and density functional tight-binding (DFTB) theory (see Section \ref{Methods}). The constructed ML models are validated by analyzing learning curves derived from the evaluation of prediction errors on a distinct molecular trajectory. In particular, the work first discusses how the prediction error decreases for a growing number of training samples on the most accurate but also numerically most expensive target fidelity level. 
As a second analysis, the actual reduction in computation time is quantified. Depending on the application, the numerical results in Section \ref{Results} show a drastic numerical gain in computational efficiency by over a factor of 30 achieved by the current method compared to classical single-fidelity KRR models. This outcome clearly shows that the present approach is a viable choice for much more complex systems and larger data sets.

\section{Results and Discussion}\label{Results}

In MFML, cost-efficient models for a given target fidelity of the excited state energy are built. For a large part of this study, the fidelities are given by different basis sets for the excited states using the TD-DFT approach with the CAM-B3LYP functional (see Section \ref{Methods}).
Therefore, the different fidelities are simply named after the basis set (or even a shorthand version thereof). 
Single-fidelity machine learning for the most accurate target fidelity, $F$, with def2-TZVP basis set hence leads to a model, which is denoted by $P_{\rm KRR}^{\rm (TZVP)}$. Please note that the accuracy of the data increases with the fidelity $f$, i.e., $f=1$ denotes the least accurate  and $f=F$ the most accurate data. 
The MFML approach replaces the model $P_{\rm KRR}^{\rm (TZVP)}$ by a cheaper-to-train model that still targets the same fidelity but contains data from a sequence of fidelities starting from the target fidelity (TZVP) down to a  \textit{baseline fidelity}, $f_b$, e.g., 3-21G. Mathematically, this is realized by first constructing a single-fidelity model on the level of the baseline fidelity and adding up several $\Delta$-ML type intermediate models for  $f_b\leq f<F$, $P_{\rm KRR}^{(f,f+1)}$ between fidelities, e.g., 6-31G, def2-SVP, leading to
$$
    P^{\footnotesize\mbox{(TZVP;3-21G)}}_{\rm MFML} := P^{\footnotesize\mbox{(3-21G)}}_{\rm KRR}
     + P_{\rm KRR}^{\footnotesize\mbox{(3-21G,6-31G)}}
      + P_{\rm KRR}^{\footnotesize\mbox{(6-31G,SVP)}}
       + P_{\rm KRR}^{\footnotesize\mbox{(SVP,TZVP)}}\,.
    \label{eq_MFML_specific}
$$
Starting from the target fidelity, the $\Delta$-ML type models are trained with an increasing amount of training samples for decreasing fidelity. Since much less training samples are used on the costly fidelities, the method becomes efficient. 
The hierarchy of fidelities to be used in MFML should have a decreasing amount of systematic error towards the target fidelity (see Sections~\ref{prelim_analysis} and \ref{Methods_MFML}).

Typically, the prediction errors for the three studied molecules benzene, naphthalene, and anthracene are discussed via learning curves (see Section~\ref{Methods_MFML}). The main objective of this analysis is to show, how the additional cheaper fidelities enhance the prediction. Improvements are reported with respect to the number of the most expensive training samples (see Section~\ref{MFML_results}) and with respect to the projected total time to generate the training data (see Section~\ref{cost_reduction_results}). The learning curves also indicate how additional training data provides a better prediction accuracy.

Since all calculations were performed along both MD and DFTB trajectories, the main manuscript in most cases shows only the results arising from the MD trajectories of the molecules.
For each trajectory, the training is performed with the data set structured as follows: on the target fidelity, that is TZVP, $1.5\cdot 2^9 = 768$ excited state energies are determined. The factor of 1.5 ensures that the training data is sufficiently different for each random shuffling needed in the model evaluation. For each subsequent lower fidelity, this number is scaled by a factor of 2 thus resulting in $1.5\cdot 2^{13}=12288$ excited state energy calculations at the lowest fidelity, that is STO-3G.
The evaluation of the MFML models are performed on a separate holdout set, also called the evaluation set, with energies calculated at the TZVP fidelity (see Section \ref{QC_data}). These conformations from the evaluation set are never used in the training of the model.
The plots for the results along the DFTB trajectories are shown in the SI. Some final results shown here, however, include results of both MD and DFTB trajectories as evidence of the method implementation across trajectory types.

\subsection{Preliminary Data Analysis}\label{prelim_analysis}

\begin{figure}[h!tb]
    \centering
    \includegraphics[width=\textwidth]{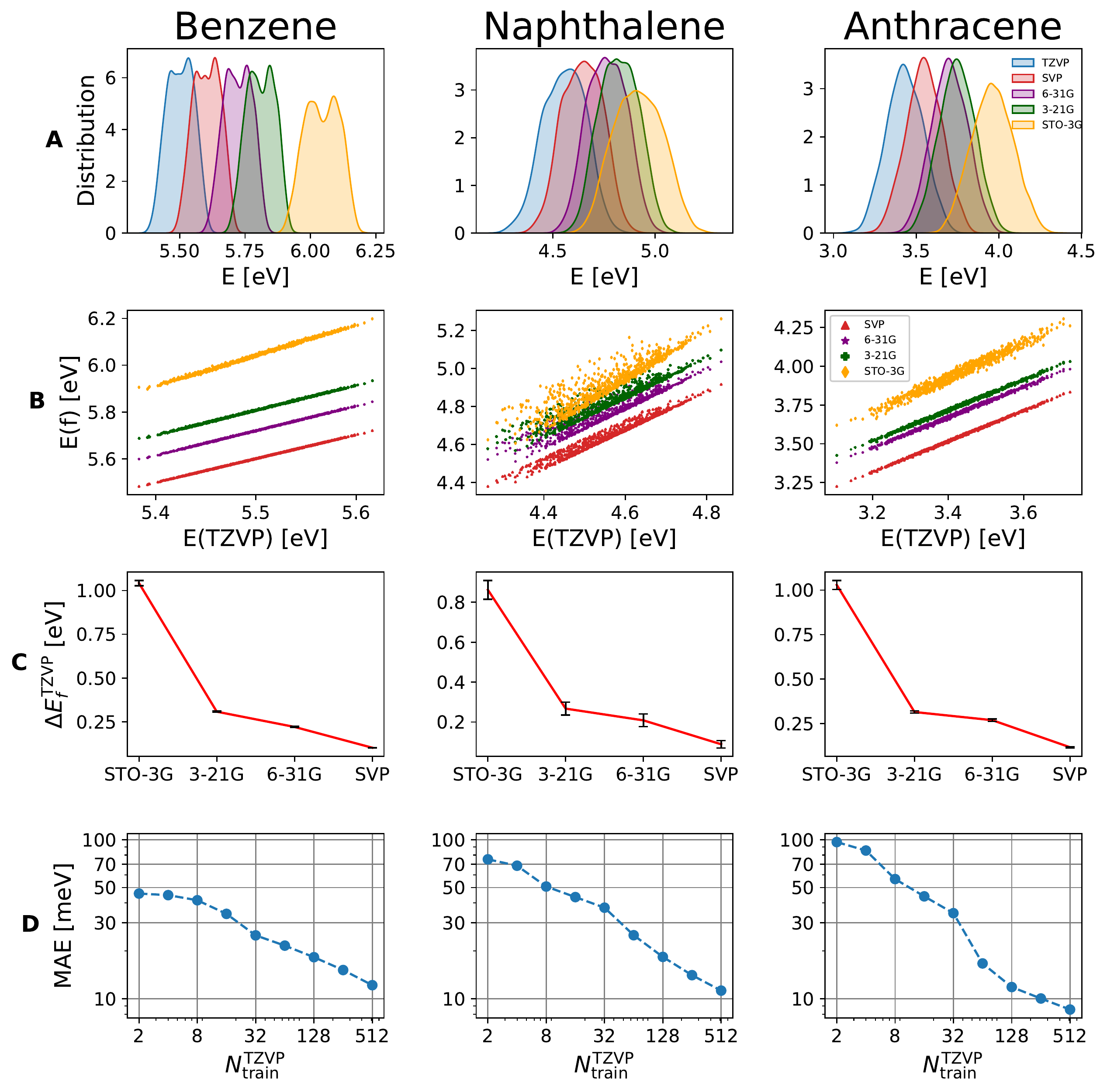}
    \caption{A) Energy distributions of the different fidelities (basis sets) in the training sets based on the MD trajectories of benzene, naphthalene, and anthracene. To aid visibility, the STO-3G distributions (yellow) have been shifted down in energy by 0.5 eV for all molecules. The complete training data for each fidelity is represented in terms of the density plot obtained using the kernel density estimation. B) Scatter plots comparing the excitation energies using the TZVP basis set to the excitation energies at the other fidelities (basis sets) for the conformations in the training data. Again, the STO-3G distributions (yellow) have been shifted down in energy by 0.5 eV. C) Energy differences (including standard deviations) between the different fidelities and the target fidelity (TZVP) for the conformations in the training data. A hierarchy in the accuracy of the different excited state calculations is a necessary condition for the working of the MFML approach. D) Learning curves for the single-fidelity KRR model presented on a double-logarithmic scale. }
    \label{fig_prelim}
\end{figure}

Before using the training data for the MFML approach, some of its characteristics need to be analyzed to verify that a clear hierarchy in the accuracy of the excited state methods really does exist. Shown in Fig.~\ref{fig_prelim}A are the energy distributions, i.e., the kernel density plots of the energy values, at different fidelities (basis sets) for the three molecules based on the classical MD trajectory. The equivalent results for the DFTB trajectory are shown in Fig.~S2A. To enhance the visual representation of the data, the excited state energies for the STO-3G basis set have been shifted by 0.5 eV towards lower energies. In several cases, the energy distributions have a Gaussian shape, while in some cases, such as the MD based benzene and DFTB based anthracene, the distributions are bimodal. This unusual bimodal shape of some distributions might be due to the finite number of samples and could be the result of a limited coverage of the conformation space arising from the use of a single trajectory.

The second type of plots shown in Fig.~\ref{fig_prelim}B are  scatter plots between the target fidelity TZVP and the other fidelities, which are present in the training data. For this stage of the analysis, only those molecular conformations were considered which belong to the training set $\mathcal{X}^{\rm TZVP}$ (see Section \ref{Methods_MFML}). The conformations from the evaluation set are not considered. This plot helps to understand how the fidelities included in the MFML model deviate from the target fidelity. For the approach to work, one anticipates that the lower fidelities have a systematic difference in accuracy to the target fidelity. In the data based on the MD but also the DFTB trajectories of benzene, the points are closely packed for each fidelity and show a nearly linear dependence between the excited state energies. The same is observed for the data based on the DFTB trajectory for naphthalene (see Fig.~S2B) and the MD trajectory for anthracene. For the MD-based naphthalene data, the SVP and 6-31G points are relatively close to a line, while the 3-21G and STO-3G results show a much larger spread.
The same is the case for the DFTB-based anthracene excitation energies determined using the STO-3G basis set when plotted against the TZVP energies as can be seen in the third frame of Fig.~S2B. Thus, not always the same amount of improvement seems to be present when increasing the basis set size for these cases. For certain molecular conformations, the increase in accuracy is larger than for others. This effect might have to do with the ability to describe the ground and/or excited state molecular orbitals with small basis sets better for some conformations than for others. 
The relatively large spread in the relationship between the target fidelity and some other fidelities
is a first hint that for some combinations of trajectory and basis set, the hierarchy in the accuracies of the different fidelities might be slightly problematic. 

To analyze this further, a test was performed to check whether the hierarchy in the accuracies of the fidelities is reflected in the training data.
For the MFML approach to yield accurate results, the method in its current version relies on the assumption that there is a systematic decrease in the calculation errors for all fidelities with respect to the target fidelity. In other words, the data used for training is expected to have a clearly decreasing error within the hierarchy. 
The mean absolute differences $\Delta E^{\rm TZVP}_{f} = \left( \sum_{i=1}^{N^{\rm TZVP}}\left\lvert E_i^{\rm TZVP}-E_i^f\right\rvert \right) /N^{\rm TZVP}$ were calculated to verify this hierarchy.
In Fig.~\ref{fig_prelim}C this quantity is used as the vertical axis and
the results are shown as a function of the different fidelities while the error bars correspond to the standard deviations of the absolute differences.
For the trajectories of benzene and anthracene, the stipulated error hierarchy requirement is satisfied, and the standard deviations do not overlap. The results for naphthalene along the DFTB trajectory and for anthracene along the MD trajectory also satisfy this requirement and show a clear hierarchy of the differences. In case of the excited state energies for naphthalene based on the MD data, it can, however, be observed that there is an overlap of the error bars for $\Delta E^{\rm TZVP}_{\rm 3-21G}$ and $\Delta E^{\rm TZVP}_{\rm 6-31G}$. This finding implies that there is no clear ordering in the accuracies of the excited state calculations for naphthalene using the 3-21G and the 6-31G basis sets, i.e., the assumed hierarchy is not clearly fulfilled for this case. This discrepancy disrupts the MFML assumptions, and it can be anticipated that the models for naphthalene along the classical MD trajectory which contain the 3-21G fidelity will be affected. 

As a final part of the preliminary analysis, the learning curves for single-fidelity KRR models, that is $P^{\rm (TZVP)}_{\rm KRR}$, are reported. The models are built on the training sets using only the TZVP fidelity and target the same fidelity. Learning curves, averaged over ten randomly shuffled training sets (see Section \ref{Methods}) have been generated for these models and are presented in Fig.~\ref{fig_prelim}D. It can be seen that these learning curves decay algebraically regardless of the molecule or on which ground state trajectory the excited state results are based. For the DFTB-based benzene, a range of low improvement for smaller training set sizes is observed. This is, however, the pre-asymptotic region and for larger training set sizes the learning curve clearly depicts a reduction in the MAE. For the number of training samples, $N_{\rm train}^{\rm TZVP}=512$, the model for the DFTB-based benzene (see Fig.~S2D) reached an MAE comparable to that of the benzene data based on the MD conformations. With 512 training samples at the TZVP level, the models for all molecules reached an MAE of the order of 10 meV. The negative slopes of the learning curves for large $N_{\rm train}^{\rm TZVP}$ indicate that further addition of training samples can potentially improve the accuracy of the predictions even further. 

\subsection{Multi-Fidelity Results}\label{MFML_results}

\begin{figure}
    \centering
    \includegraphics[width=\textwidth]{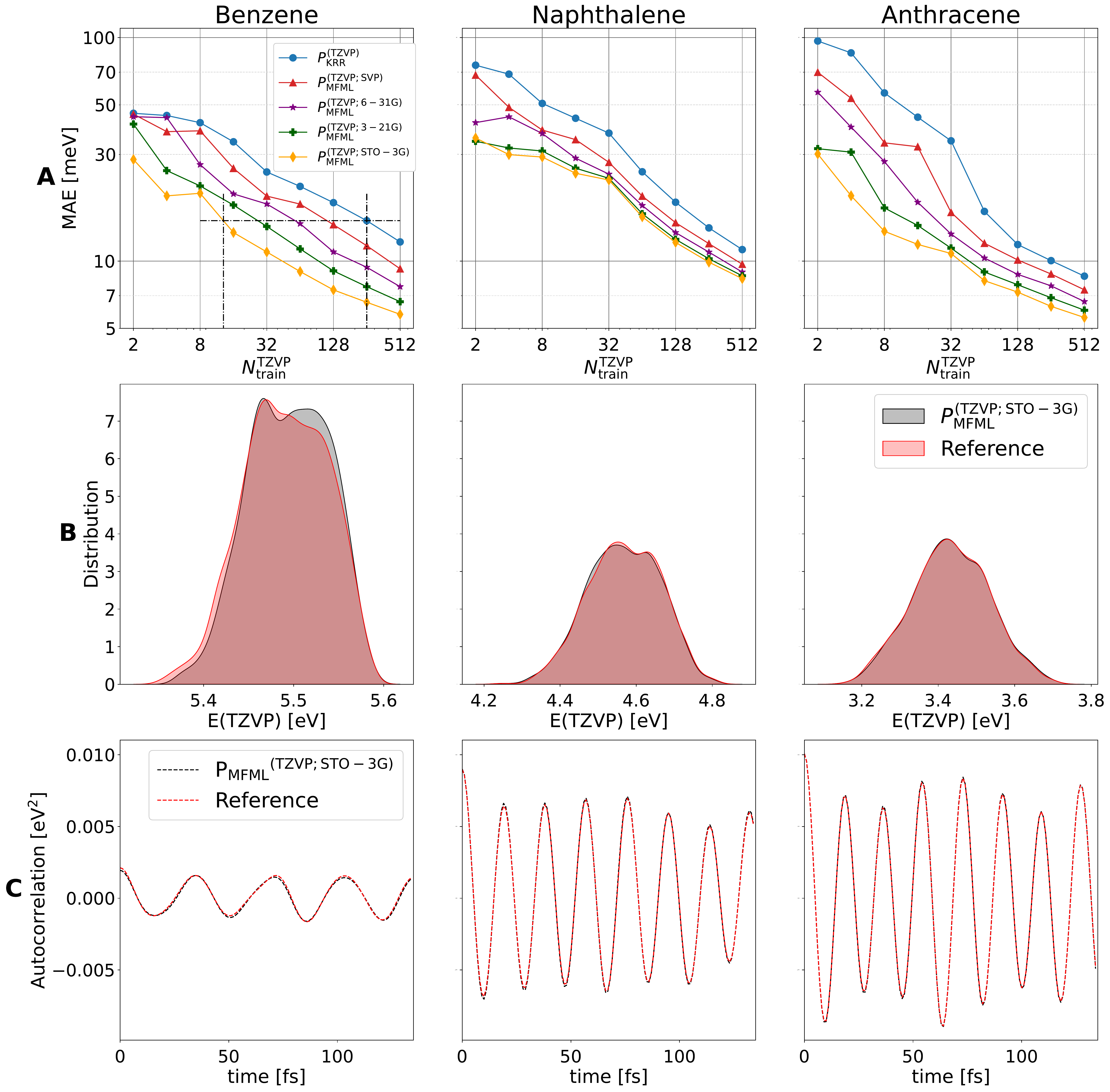}
    \caption{The effectiveness of the MFML method is represented through learning curves, while the results for the evaluation set are also analyzed 
    in the time and energy domains. A) Multi-fidelity learning curves based on the excited state energies along the MD trajectories for benzene, naphthalene, and anthracene. With the addition of lower fidelities, the prediction error decreases, as can be seen in the difference between the standard KRR model (blue) and the MFML model using data from all five fidelities (yellow). 
    B) Energy distributions based on the holdout sets using the TZVP reference calculations (red) and the predictions from the MFML model $P_{\rm MFML}^{(\rm TZVP;STO-3G)}$ for $N_{\rm train}^{\rm TZVP}=512$ (black). In all cases, it can be observed that the predictions from the MFML model matches the reference energy distributions accurately. C) The corresponding time autocorrelation functions (ACFs) of the excited state energies. The red lines correspond to the ACFs of the TZVP reference calculations from the holdout set, while the black lines report the ACF of the excited state energies predicted from the MFML model for the conformations belonging to this set.}
    \label{fig_MF_LC}
\end{figure}

Continuing with the present study, the results of the MFML approach are shown and discussed next.
In Fig.~\ref{fig_MF_LC}A the MFML learning curves for the MD trajectories are delineated. The top blue line in each of these panels refers to the standard single-fidelity KRR model, as already shown in Fig.~\ref{fig_prelim}D. The other lines correspond to MFML models built using an increasing number of fidelities in the models. In detail, this means that the first MFML model includes the
SVP data ($f=4$) in addition to TZVP. The subsequent models each include an additional fidelity, i.e., 6-31G ($f=3$), 3-21G ($f=2$), and STO-3G ($f=1$). Thus, the most elaborate MFML model includes data from five different excited state calculations, where the number of data points increases by a factor of two when going down in the fidelity.
Horizontal and vertical dashed lines were included in Fig.~\ref{fig_MF_LC}A for the case of benzene to highlight that the addition of cheaper fidelities does in fact reduce the MAE for a given training set size at the target fidelity. The horizontal line is drawn at the MAE value resulting from the single fidelity standard KRR model $P^{\rm(TZVP)}_{\rm KRR}$ at $N_{\rm train}^{\rm TZVP}=256$.
At the position where this line intersects with the line for the MFML model $P^{\rm(TZVP;STO-3G)}_{\rm MFML}$, the left vertical dashed line is depicted. The horizontal position of this perpendicular line corresponds to a value of about 16 training samples at the TZVP fidelity. Thus, the error for the MFML model $P^{\rm(TZVP;STO-3G)}_{\rm MFML}$ with about 16 training samples at the TZVP level is roughly the same as the one for the single fidelity model 
$P^{\rm(TZVP)}_{\rm KRR}$ with 256 data points at the same theory level. Certainly, for the MFML approach, calculations at the other fidelities were involved, i.e., the model is built with the number of training samples $N^f=(2^8,2^7,2^6,2^5,2^4)=(256,128,64,32,16)$ at the fidelities $f=(1,2,3,4,5)$ which are ordered as explained in Section \ref{Methods_MFML}. This finding shows a significant reduction in the number of samples in the numerically costly training data set required to achieve a certain MAE of prediction. In this specific example of the benzene molecule and an MD trajectory using a value of 512 for $N_{\rm train}^{\rm TZVP}$, the MAE were 12.2 meV and 5.7 meV for $P^{\rm (TZVP;STO-3G)}_{\rm MFML}$ and $P^{\rm (TZVP)}_{\rm KRR}$, respectively.

The learning curves for the excited states of naphthalene along the MD trajectory also show a clear and systematic offset between the standard KRR model and the MFML models, as can be seen in Fig.~\ref{fig_MF_LC}A. The addition of the STO-3G fidelity and to some extent of the 3-21G basis set, however, did not improve the model significantly. Despite the offsets between the learning curves, the MAE values for $P^{\rm (TZVP;3-21G)}_{\rm MFML}$ and $P^{\rm (TZVP;STO-3G)}_{\rm MFML}$ are very similar, i.e., 8.6 meV and 8.4 meV, respectively. 
Already in Section \ref{prelim_analysis}, based on the lack of a clear hierarchy of methods in Fig.~\ref{fig_prelim}C, it was anticipated that the MFML scheme might not provide perfect results. From the learning curves, it can now be seen that the method did actually work for the present case. However, the improvement for some fidelity levels was only marginal, while including most of the other fidelity levels did result in an increase in accuracy. For $N_{\rm train}^{\rm TZVP}=512$ the standard KRR model $P_{\rm KRR}^{\rm (TZVP)}$ yields an MAE of 11.2 meV while $P_{\rm MFML}^{\rm (TZVP;SVP)}$ and $P_{\rm MFML}^{\rm (TZVP;6-31G)}$ reach smaller MAE values of 9.6 meV and 8.9 meV, respectively. It is worthwhile to mention that in spite of the irregularities in the data, the multi-fidelity model $P_{\rm MFML}^{\rm (TZVP;STO-3G)}$ still results in lower error values than the preceding models. This finding again indicates the robustness of the present MFML method.

For the MD trajectory of anthracene, the learning curves are reported in Fig.~\ref{fig_MF_LC}A as well. The addition of each cheaper fidelity shows a clear and distinct reduction in the MAE indicating the effectiveness of the approach. For $N_{\rm train}^{\rm TZVP}=512$, the averaged MAE for the standard KRR model, $P_{\rm KRR}^{\rm (TZVP)}$, was 8.6 meV. In comparison, the multi-fidelity model, $P_{\rm MFML}^{\rm (TZVP;STO-3G)}$ resulted in an averaged MAE of 5.5 meV. In addition, Fig.~S3A shows the MFML learning curves for the DFTB-based trajectories of the various molecules. For benzene, these show a trend similar to the MD trajectory results. The averaged MAE for $P_{KRR}^{(TZVP)}$ and $P_{MFML}^{(TZVP;STO-3G)}$ were 10.8 meV and 6.7 meV, respectively.
In the learning curves for DFTB-based naphthalene shown in Fig.~S3A, the MFML models built with each additional less accurate fidelity, show lower offsets for various training set sizes. That is, if one considers a vertical line drawn at some $N_{train}^{TZVP}$, then the learning curves with the less accurate fidelities fall below the learning curves of the preceding models. There is a jump observed for all learning curves between $N_{train}^{TZVP}\approx 16$ and $\approx 32$, which is carried forward due to the jump observed in the conventional KRR model. The subsequent multi-fidelity models were built including the TZVP data, and thus this jump is also included in their results. For $N_{train}^{TZVP}=512$, for example, the averaged MAE for $P_{KRR}^{(TZVP)}$ is 8.9 meV and for $P_{MFML}^{(TZVP;STO-3G)}$ 6.5 meV. Therefore, the addition of training samples from the less accurate but numerically cheaper fidelities does in fact reduce the error of the models built for DFTB naphthalene.

While the DFTB-based anthracene showed conformation to the energy difference hierarchy as seen in Fig.~S2C, it can be observed that the MFML model does not provide an improvement with the addition of STO-3G fidelity. This is shown in the MFML learning curves on the right-hand side of Fig.~S3A.
For smaller training set sizes, the learning curve corresponding to the model $P^{(TZVP;STO-3G)}_{MFML}$ crosses above that of the MFML model built on the 3-21G baseline. However, for larger training set sizes, the robustness of the MFML approach succeeds, resulting in a comparable MAE for $P_{MFML}^{(TZVP;STO-3G)}$ as can be seen for $N_{train}^{TZVP}=512$. The details of this specific case are further elaborated in Section S3.1.

\subsection{Predictions in the Energy and Time Domains}

Before analyzing the computational costs of the MFML scheme, the results of this approach are analyzed in ways different from learning curves and slightly closer to applications, e.g., in the calculation of spectral densities \cite{mait23a}. Such properties might not really be relevant for molecules in the gas phase, but will become essential once similar calculations will be performed for molecules in non-trivial environments. To this end, Fig.~\ref{fig_MF_LC}B compares the distributions of the excited states along the MD trajectory. This comparison is performed between the TZVP reference energies from the holdout set with those predicted for the conformations in this evaluation set by the MFML formalism using five fidelities, i.e., the $P_{\rm MFML}^{\rm (TZVP;STO-3G)}$ model for $N_{\rm train}^{\rm TZVP}=512$.
A visual comparison yields basically no differences for the molecules naphthalene and anthracene, while for benzene, small differences in the peak structure are visible. For most applications, this level of accuracy is certainly more than necessary. 
Looking again at only the training data at the different fidelities for benzene in Fig.~\ref{fig_prelim}A, it becomes evident that the training data has a bimodal distribution, which translates to the $P_{\rm MFML}^{\rm (TZVP;STO-3G)}$ model. If the models were to be trained on a larger dataset, this likely would be smoothed out since a larger section of the conformation space would be covered and the bimodality in the training set most likely would disappear, being an artifact of the small number of training data along a trajectory.
A similar agreement is observed for the MFML model for the DFTB-based trajectories of the molecules, as can be seen in Fig.~S3B.

After having analyzed the data in the energy domain, the next step was to have a closer look at the time domain. Instead of looking at individual arbitrary pieces of the trajectory, the autocorrelation function (ACF), which can also be averaged in a meaningful way, was analyzed. The ACF for a discrete time series can be determined as \cite{damj02a}
\begin{equation}
C_m(t_l) = \frac{1}{N-l} \sum_{k=1}^{N-l} \Delta E_m(t_l + t_k) \Delta E_m(t_k)\,,
\label{eq:acf}
\end{equation} 
where $\Delta E_m$ denotes the difference between the excitation energies $E_m$ and the time average $\langle E_m \rangle$, i.e., $\Delta E_m(t) = E_m(t) -\langle E_m \rangle$. 
Moreover, $N$ represents the number of frames present in the respective part of the trajectory. The initial 2700 frames from the evaluation data set were taken into account for each molecule. These trajectories were divided into ten independent windows, each with 270 conformations. Since a time step of 1~fs was employed, an ACF of a length of 135~fs was constructed using this data. The correlation functions were averaged over the ten windows. 
In Fig.~\ref{fig_MF_LC}C the reference data, i.e., excitation energies along the MD trajectory determined using the TZVP fidelity, is compared to the predictions from the $P_{\rm MFML}^{\rm (TZVP;STO-3G)}$ model built with $N_{\rm train}^{\rm TZVP} = 512$.
It is clearly visible that the predictions from the MFML model in the case of this averaged ACF reproduce the results obtained at the TZVP level with high accuracy. In Fig.~S3C, a similar agreement can be observed for the excitation energies along the DFTB trajectories.

\subsection{Reduction of Computation Time for Generating Training Data} \label{cost_reduction_results}

Finally, as the most important part of the present study, the decrease in the computation time needed to generate the training data when using MFML is studied. To this end, the MAE will not be studied as a function of the number of training samples on the highest fidelity, but as a function of the computation time to generate the complete training data on all hierarchy levels. The average computation times of single point calculations at the different fidelities performed are reported in Table S1 for the three studied molecules. Based on this data, the total time to generate the training sets for a given model can be determined as $\sum_{f=f_b}^{F}N^{(f)}\cdot \overline{T}^{(f)}$, where the sum runs from the baseline fidelity $f_b$ up to the target fidelity, $F$, which in this case is TZVP. In this expression, $N^{(f)}$ denotes the number of training samples used for fidelity $f$ and $\overline{T}^{(f)}$ denotes the corresponding average computation time for the respective single point calculation, as reported in Table S3. For example, the computational time to generate the training set for benzene to construct the model $P_{\rm MFML}^{\rm (TZVP;6-31G)}$ with $N_{\rm train}^{\rm TZVP}=4$ training samples at the target fidelity can be estimated to be $T_{\rm train}^{\rm MFML} = 4 \cdot 3.53 \text{ min} + 8 \cdot 1.02 \text{ min}+ 16 \cdot 0.65 \text{ min} \approx 32 \text{ min}$. 

\begin{figure}
    \centering
    \includegraphics[width=\textwidth]{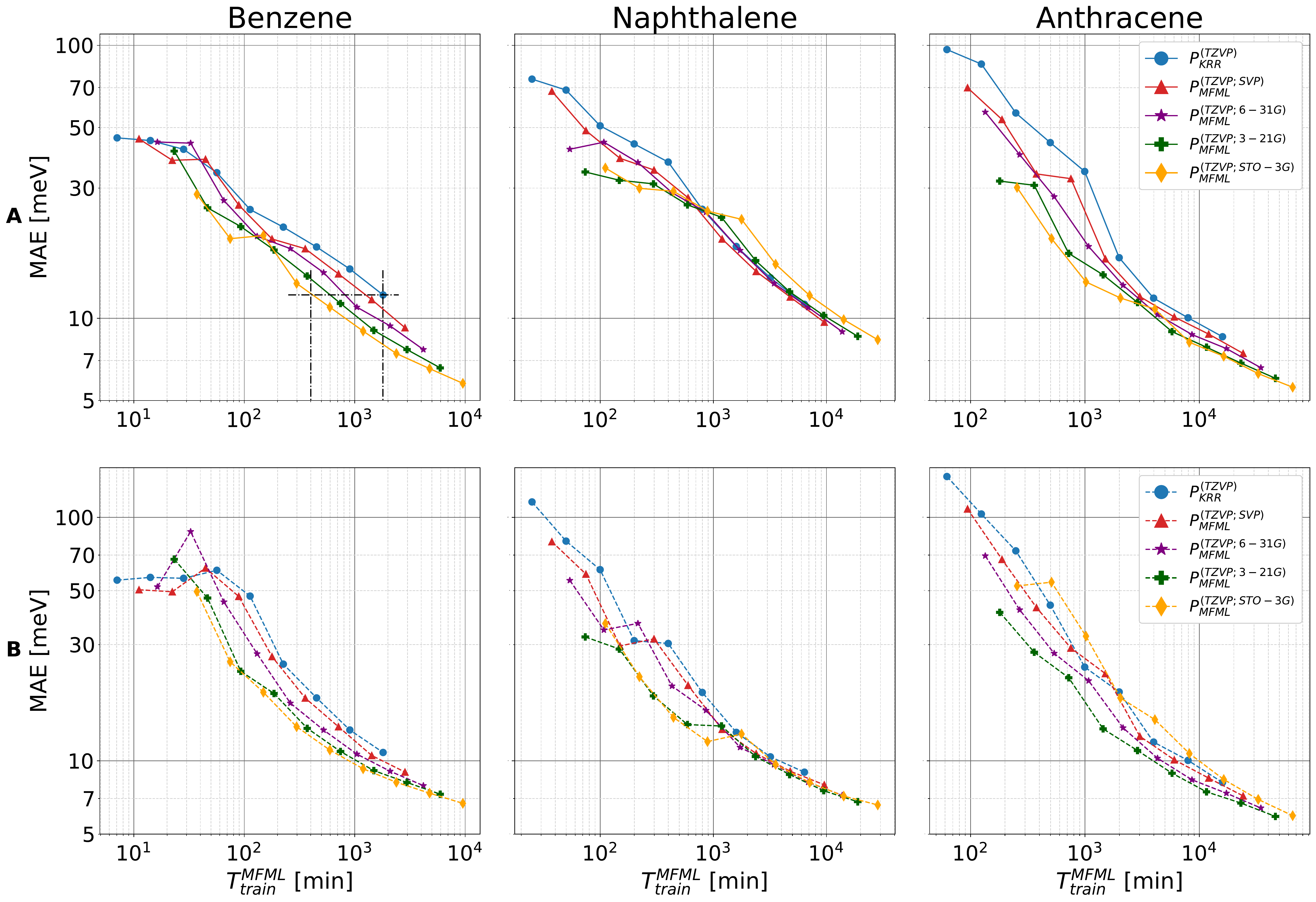}
    \caption{Computation times to generate the training data sets versus the MAE of the MFML models, verifying the computational benefits of the MFML models. 
    A) Results for the MD trajectories: With addition of each additional numerically cheaper fidelity, the training time decreases for a specific MAE, i.e., prediction accuracy, in the cases with a clear hierarchy in the fidelities.
    B) Findings for the DFTB trajectories: The time benefits are clearly visible for the various molecules across the fidelities. For anthracene, the yellow line corresponding to the MFML model built on STO-3G does not provide any time improvement due to the various aforementioned reasons.
    }
    \label{fig_time}
\end{figure}

Shown in Fig.~\ref{fig_time}A are the MAEs as a function of the
the projected computation times for generating the training data sets for conformations of the three molecules benzene, naphthalene, and anthracene along the classical MD trajectory. 
For the case of benzene, dashed lines have been included again to help to interpret the data.
For $N^{\rm TZVP}_{\rm train}=512$, the projected computation time to generate the training set only at the target fidelity is approximately 1800 minutes. Using this data, one can generate the standard KRR model $P^{\rm (TZVP)}_{\rm KRR}$. The horizontal line shows that a similar MAE can be achieved for the MFML model $P_{\rm MFML}^{\rm (TZVP;STO-3G)}$ for which the projected time to generate the training set is about 400 minutes. Thus, the MFML method provides a factor of roughly 4.5 in reducing the computation time required for the generation of the training set. 
The plot of time to generate the training data set versus the MAE for the MD-based naphthalene conformations in Fig.~\ref{fig_time}A shows again that the MFML model is affected if the difference hierarchy structure is not upheld for the training data as explained in Section \ref{prelim_analysis}. Thus, it becomes all the more important to ensure that the employed training data follows the assumed hierarchy of basis set sizes or quantum chemistry methods. While the MFML approach still remains robust, the cost of the training data generation might not always follow suit unless the assumed hierarchy holds throughout the training data.
In addition, the time to generate the training data set versus MAE for anthracene based on the MD trajectory given in Fig.~\ref{fig_time}A reflects the results of the corresponding learning curves in Fig.~\ref{fig_MF_LC}A. A computational cost reduction in the training data generation time is observed across the multi-fidelity models. The standard KRR model $P^{\rm (TZVP)}_{\rm KRR}$ at $N_{\rm train}^{\rm TZVP}=512$ yields a projected time of about 16000 minutes while the multi-fidelity model $P_{\rm MFML}^{\rm (TZVP;STO-3G)}$ gives a similar error with a training set generation time of roughly 7000 minutes, which results in a cost reduction by a factor of about 2.3 across the multi-fidelity model. The model $P_{\rm MFML}^{(TZVP;3-21G)}$ results in a similar error for a training set with a projected time of generation roughly 6000 minutes, which corresponds to a time benefit factor of about 2.7 for the training data generation cost.

Similarly, for the molecules based on the DFTB trajectory, the time to generate the training data sets versus the error in prediction is shown in Fig.~\ref{fig_time}B. For benzene, the benefit of the MFML is evident. If one draws reference lines again for this plot, one observes that the projected time to generate the training set for benzene to train the model $P^{\rm (TZVP)}_{\rm KRR}$ is about 1800 minutes, whereas the time to generate the training set for $P^{\rm (TZVP;STO-3G)}_{\rm MFML}$ to result in a similar MAE is roughly 600. This corresponds to a saving in the computational time by a factor of 3.

For DFTB-based naphthalene in Fig.~\ref{fig_time}B, one observes that in the case of $N^{\rm TZVP}_{\rm train} = 32$ a jump occurs for the $P^{\rm (TZVP;STO-3G)}_{\rm MFML}$ model. This jump occurs due to a jump which is already present in the MAE values for $P^{\rm (TZVP)}_{\rm KRR}$ as shown in Fig.~S2D. It can be understood, since the TZVP data is contained in both models. One has to notice, however, that the computational time to generate $512$ TZVP training samples is close to 6400 minutes. The $P^{\rm (TZVP;STO-3G)}_{\rm MFML}$ model results in a similar error for a computational time for generating the training data set of about 3000 minutes. The MFML model thus reduces the time cost for the generation of the training set by a factor of about 2 for DFTB-based naphthalene.

As can be seen in Fig.~\ref{fig_time}B, for the DFTB-based trajectory of anthracene, the MFML model $P^{\rm(TZVP;STO-3G)}_{\rm MFML}$ performs poorly and therefore does not
provide any cost reduction. 
As discussed in Sections \ref{prelim_analysis} and \ref{Methods}, this is due to the scattered nature of the STO-3G data, which is depicted in Fig.~S2B. The effect of such a wide scatter on the difference models is further explained in the SI.
For the other MFML models of the same system, a reduction in computational training time is still clearly visible. The time to generate $512$ training samples at the TZVP fidelity is about 16000 minutes. The prediction error achieved by $P^{\rm(TZVP)}_{\rm KRR}$ for this training size can be achieved by the $P^{\rm(TZV P;3-21G)}_{\rm MFML}$ MFML model for a training data set with a computational time of roughly 7000 minutes. This corresponds to cost reduction by a factor of 2.3 in the training data generation time resulting while achieving a similar accuracy. 
This example once more shows the importance of the existence of a clear hierarchy of fidelity levels, and that overlapping regions of accuracy for different schemes can hinder an improvement of the results in a MFML approach.

\subsubsection{Additional levels of fidelity}

\begin{figure}[!hbt]
    \centering
    \includegraphics[width=\textwidth]{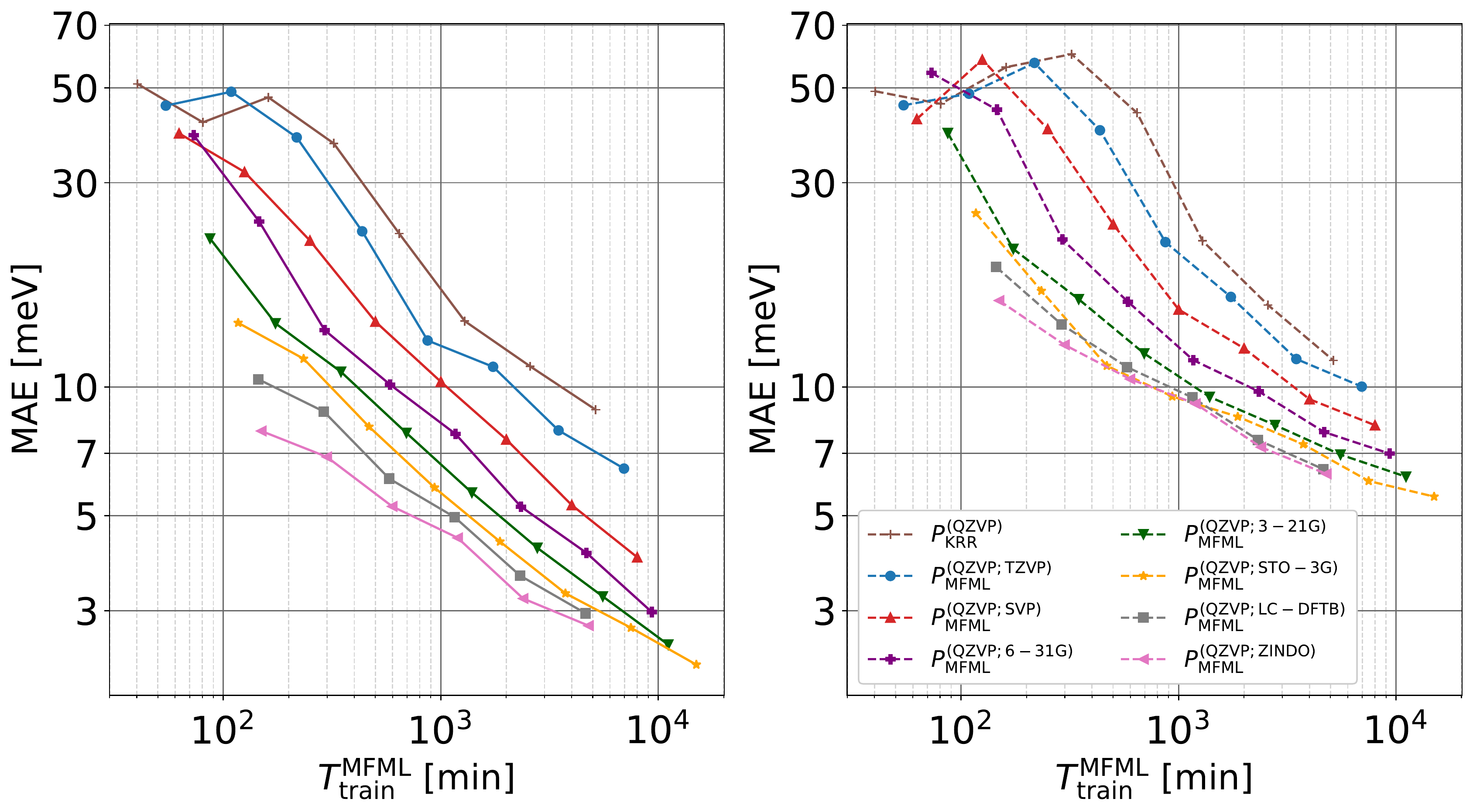}
    \caption{Computational time to generate the MFML training set versus the MAE for benzene. The results for the MD-based trajectory are presented on the left-hand side, while the right-hand side shows the results for the DFTB-based trajectory. The target fidelity is set to QZVP. Additionally, two semi-empirical methods, ZINDO and LC-DFTB were employed. For each numerically cheaper fidelity that is added into the model, clear offsets of the learning curves can be observed. }
    \label{fig_ben_qzvp}
\end{figure}
Calculating excited state energies for naphthalene and anthracene along trajectories using TDDFT with basis sets larger than TZVP, becomes numerically quite expensive. Thus, the analysis was only furthered for the two trajectories of benzene by training the MFML model to predict the first excited state energy using the def2-QZVP basis set for the TD-DFT calculations, which is considerably larger than the def2-TZVP. Thus, it is numerically costlier to calculate the first excited state energy for this fidelity. In addition to these more accurate calculations for the excited states of benzene, two semi-empirical methods, namely, LC-DFTB and ZINDO were used. It is assumed that ZINDO is the least accurate approach, followed by LC-DFTB and then TD-DFT with the CAM-B3LYP functional and the diverse basis sets as studied above. Fig.~\ref{fig_ben_qzvp} shows the plots for the computational time for the generation of the training data versus the MAE for benzene along the MD and the DFTB trajectories. In this case, the models are built to target the numerically most expensive QZVP fidelity. 

First, we consider the plot for MD-based benzene. The standard KRR model $P_{\rm KRR}^{\rm(QZVP)}$, with 256 training samples at QZVP fails to reach the accuracy offered by $P_{\rm MFML}^{\rm(QZVP;ZINDO)}$ with merely 2 training samples at QZVP. On comparing the time required to generate the training set, one observes that the standard KRR model required about 5000 minutes while the MFML model achieves a lower MAE for about 150 minutes. This represents a time benefit of over a factor of 30. 
If the trend of the MAE for the KRR model were to continue, i.e., if the curve for $P_{\rm KRR}^{\rm(QZVP)}$ would be slightly extended, we extrapolate a time benefit of over 50 while using the MFML model.

For the benzene conformations along the DFTB trajectory, although the MAEs are slightly larger than those for the data based on the MD trajectory, the time benefit is about a factor of 17 with roughly 5000 versus 300 minutes for the $P_{\rm KRR}^{\rm(QZVP)}$ and the $P_{\rm MFML}^{\rm(QZVP;ZINDO)}$ models, respectively.
Even the MFML model $P_{\rm MFML}^{\rm(QZVP;STO-3G)}$ yields a computational time benefit of larger than 10-fold for the DFTB and MD trajectories of benzene. Thus, it becomes evident that for the presented MFML approach, the savings in computational time for the training data set generation tend to be larger when the most accurate fidelity is numerically much more expensive than the lower fidelity levels.

\section{Conclusion}\label{Conclusion}

This work has introduced the utilization of MFML models for the prediction of excited states, here with a focus on the first excited state. With molecules of various sizes, it has been shown that if the hierarchy of the fidelities holds, this method does in fact reduce the computation time for training the models while improving the accuracy of the predictions. 
The results and methods from this work deliver impetus in the direction of showing that the machine learning methods can be made more efficient by the use of data in a multi-fidelity structure. While in this work, the method is applied to the first excited state energy, the overall method can be applied to any property where a hierarchy of training data can be established.
MFML is not restricted to the dimension of the basis set and can be generalized to multiple dimensions provided the hierarchy holds \cite{zasp19a}. It is recommended to perform a preliminary data analysis to ensure that the hierarchy of the method holds while generating the training set. 
Future work should extend the method towards being applicable to less clearly defined hierarchies of methods.
Optimizing the factor that scales the number of data points between the fidelities is another interesting point for future research in MFML. Here, one should systematically assess the effect of the ratio $N_{f+1}/{N_f}$ for all $f_b\leq f<F$ on the prediction errors. Understanding this relationship can potentially lead to an approach that further reduces the time required to generate the training sets. 

Overall, this work has numerically shown that for unseen data, the MFML method can predict the first excited state energy with a high level of accuracy as made evident by the learning curves. The method also achieves to maintain the distributions of energies and their time correlations along dynamical trajectories, thereby being a strong contender to high-accuracy low-cost machine learning models for excited state properties. Specifically, for the first excited state energy, this method has achieved a time reduction by a factor of 30 and more. In case one wants to achieve highly accurate excitation energy gaps by electronic structure methods which might scale with higher powers in the number of atoms such as coupled cluster theories, the MFML approach very likely will lead to even much larger numerical gain factors. Combining this with the wish to do such calculations along trajectories, e.g., of a chromatophore with more than 2000 pigment molecules, gives an idea how large the reduction in numerical cost reduction might become. The same is true if one wants to determine excited state potential energy surfaces or perform non-adiabatic dynamics.

\section{Methods}\label{Methods}

\subsection{Quantum Chemistry Calculations}\label{QC_data}

The training data sets for the excitation energies were generated in gas phase for the three molecules benzene, naphthalene, and anthracene along classical MD and DFTB trajectories mainly based on TD-DFT calculations with various basis sets. In the case of the classical MD simulations, the GAFF force field prepared by the ACEPYPE interface \cite{dasi12a} and the GROMACS-2022.3 package \cite{abra15a} were employed to perform the molecular dynamics simulations. First, an energy minimization was performed followed by a 100~ps-long equilibration at 300 K. Subsequently, a 100~ps run was performed. Finally, a 15~ps-long unbiased NPT simulation was carried in which the geometries were stored at every time step, i.e., every 1~fs. This procedure yielded a total of 15,000 frames, which were then utilized for excited state calculations to be used as training as well as evaluation data sets.
In case of the DFTB simulations, the 3OB parameter set \cite{gaus13a} was employed as chosen in the DFTB+ package version 21.1 \cite{hour20a} and 15~ps-long NVT simulations were carried out for the three molecules. Again, the trajectories were stored using a 1~fs stride, producing 15,000 frames for the excited state calculations. Subsequently, the first excited states of the molecules were determined along the trajectories using the TD-DFT formalism with the CAM-B3LYP functional as implemented in the ORCA package version 4.1.2 \cite{nees18a}. In all cases, five different basis sets according to their hierarchy were employed, i.e., STO-3G, 3-21G, 6-31G, def2-SVP and def2-TZVP. The exact hierarchy of excited state calculations across the 15,000 frames is discussed in the SI. In case of benzene, the larger basis set def2-QZVP was tested as well. During these calculations, the Tamm-Dancoff approximation (TDA) approximation was employed together with the Resolution of Identity approximation (RIJCOSX) in order to speed up the calculations. In addition, the computationally cheap semi-empirical methods ZINDO/S-CIS(10,10) (Zerner’s intermediate neglect of differential orbital method with spectroscopic parameters together with configuration interaction using single excitations including the 10 highest occupied molecular orbitals (HOMOs) and the 10 lowest unoccupied molecular orbitals (LUMOs) as an active space) and TD-LC-DFTB (time-dependent long-range corrected DFTB)\cite{bold20a} were employed to determine the excitation energies of the benzene molecule along both the MD and the DFTB trajectories. In shorthand notation, ZINDO/S-CIS(10,10) and TD-LC-DFTB are described as ZINDO and LC-DFTB in the respective part of Section \ref{Results}. The ZINDO calculations were performed using the ORCA package, whereas the LC-DFTB calculations were conducted using the DFTB+ package. 

\subsection{Kernel Ridge Regression}
Let a training data set $\mathcal{T}:=\{(\boldsymbol{X}_i,E_i)\}_{i=1}^{{N}_{\rm train}}$ of size ${N}_{\rm train}$ with molecular descriptors or representations $\boldsymbol{X}_i$ and their corresponding first excited state energies $E_i$ be given.
The KRR model for the prediction of the first excited state energy for an unseen query descriptor $\boldsymbol{X}_q$ is denoted by
\begin{equation}
    P_{\rm KRR}\left(\boldsymbol{X}_q\right) := \sum_{i=1}^{N_{\rm train}} \alpha_i k\left(\boldsymbol{X}_q,\boldsymbol{X}_i\right)~,
    \label{eq_KRR_def}
\end{equation}
where $k$ is the kernel function. The unknown coefficient vector $\boldsymbol{\alpha}$ is trained by solving the linear system of equations $(\boldsymbol{K}+\lambda \boldsymbol{I}) \boldsymbol{\alpha} = \boldsymbol{E}$, with $\boldsymbol{K} = \left(k(\boldsymbol{}X_i,\boldsymbol{X}_j)\right)_{i,j=1}^{{N}_{\rm train}}$ the kernel matrix, $\boldsymbol{I}$ the identity matrix, $\boldsymbol{E} = \left(E_1, E_2, \ldots, E_{{N}_{\rm train}}\right)^T$ the vector of energies and $\lambda$ a regularization parameter. 
This work uses the Matérn Kernel of first order with the discrete L-2 norm
\begin{equation}
    k\left(\boldsymbol{X}_i,\boldsymbol{X}_j\right) = \exp{\left(-\frac{\sqrt{3}}{\sigma}\left\lVert \boldsymbol{X}_i-\boldsymbol{X}_j\right\rVert_2^2\right)}\left(1+\frac{\sqrt{3}}{\sigma}\left\lVert \boldsymbol{X}_i-\boldsymbol{X}_j\right\rVert_2^2\right)~,
\end{equation}
where $\sigma$ denotes a length scale hyperparameter that determines the width of the kernel. This property is, in some sense, a measure of the degree of correlation associated with the training samples \cite{vu2015understanding, snyder2012finding, Faber2018}. In the present study, $\sigma$ was manually converged for each molecule and trajectory, and the respective values are listed in Table S2.

\subsection{Multi-Fidelity Machine Learning}\label{Methods_MFML}
The MFML approach requires an ordered sequence of fidelities $f=1,2, \ldots, F$, where the fidelities are indexed in order of numerical expense (and usually accuracy) of related energy calculations. For each fidelity $f$, representations and energy calculations at this fidelity form the training set $\mathcal{T}^{\left(f\right)}:=\left\{(\boldsymbol{X}_i, E^f_i)\right\}_{i=1}^{N_{\rm train}^{(f)}}$, where $N_{\rm train}^{(f)}$ corresponds to the number of samples at fidelity $f$. 
Defining the set of molecular descriptors $\mathcal{X}^f = \left\{\boldsymbol{X}_i^f\lvert \left(\boldsymbol{X}_i^f,E^f_i\right)\in\mathcal{T}^{(f)}\right\}$, the presented MFML approach, further requires the nestedness $\mathcal{X}^F\subseteq \ldots \subseteq \mathcal{X}^2 \subseteq \mathcal{X}^1$ of the training data. In other words, if a molecular conformation is picked which has the first excited state energy calculated at the highest fidelity, then it is also that the first excited state energy is calculated for this conformation at the second-highest fidelity, and so on.

As one of us has previously shown \cite{zasp19a}, the MFML model for the target fidelity $F$ can be iteratively built with KRR for a baseline fidelity $f_b$ as 
\begin{equation}
    P^{(F;f_b)}_{\rm MFML}\left(\boldsymbol{X}_q\right) := P^{(f_b)}_{\rm KRR}\left(\boldsymbol{X}_q\right) + \sum_{f_b\leq f<F}P_{\rm KRR}^{(f,f+1)}\left(\boldsymbol{X}_q\right)
    \label{eq_MFML_}
\end{equation}
for $f_b=1,2,\ldots,F-1$. Here, $P_{\rm KRR}^{(f_b)}$ denotes the standard KRR model built from the baseline fidelity as mentioned in Eq.~\eqref{eq_KRR_def}. 
Furthermore, the term inside the summation is defined as
\begin{equation}
    P_{\rm KRR}^{(f,f+1)}\left(\boldsymbol{X}_q\right) := \sum_{i=1}^{N_{\rm train}^{(f+1)}}\alpha_i^{(f,f+1)} k\left(\boldsymbol{X}_i,\boldsymbol{X}_q\right)
    \label{eq_MFML_singleKRR}~. 
\end{equation}
In this equation, the coefficients $\alpha_i^{(f,f+1)}$ are calculated by solving the linear system of equations 
\begin{equation}
    \left(\boldsymbol{K}+\lambda \boldsymbol{I}_{N^{(f+1)}}\right)\boldsymbol{\alpha}^{(f,f+1)} = \boldsymbol{\Delta E}^{(f,f+1)}~,
    \label{eq_MFML_alphas}
\end{equation}
where $\boldsymbol{\Delta E}^{(f,f+1)}=\boldsymbol{E}^{f+1}-\boldsymbol{E}^{(f,f+1)}$,
with $\boldsymbol{E}^{f+1}$ being the vector of energies in training set $\mathcal{T}^{(f+1)}$ and $\boldsymbol{E}^{(f,f+1)}$ the vector of energies in training set $\mathcal{T}^{(f)}$ restricted to those conformations only found on fidelity level $f+1$. Thus, a model for a target fidelity $F=5$ with a baseline of $f_b=3$ can be iteratively built as 
\begin{equation}
    P^{(5;3)}_{\rm MFML}\left(\boldsymbol{X}_q\right) := P_{\rm KRR}^{(3)}\left(\boldsymbol{X}_q\right) + P_{\rm KRR}^{(3,4)}\left(\boldsymbol{X}_q\right) + P_{\rm KRR}^{(4,5)}\left(\boldsymbol{X}_q\right)
    \label{eq_MFML_examples}
\end{equation}
with 
\begin{subequations}
\begin{equation}
    P_{\rm KRR}^{(3)}\left(\boldsymbol{X}_q\right) := \sum_{i_3=1}^{N^{(3)}_{\rm train}}\alpha^{(3)}_{i_3}k\left(\boldsymbol{X}_{i_3},\boldsymbol{X}_q\right)
\end{equation}
\begin{equation}
    P_{\rm KRR}^{(3,4)}\left(\boldsymbol{X}_q\right) := \sum_{i_4=1}^{N^{(4)}_{\rm train}}\alpha^{(3,4)}_{i_4}k\left(\boldsymbol{X}_{i_4},\boldsymbol{X}_q\right)
\end{equation}
\begin{equation}
    P_{\rm KRR}^{(4,5)}\left(\boldsymbol{X}_q\right) := \sum_{i_5=1}^{N^{(5)}_{\rm train}}\alpha^{(4,5)}_{i_5}k\left(\boldsymbol{X}_{i_5},\boldsymbol{X}_q\right)
\end{equation}
\end{subequations}
The number of training samples used for each fidelity differs by a \textit{scaling factor} of 2 based on research on sparse grid combination methods \cite{hegland2016combination,harbrecht2013combination,reisinger2013combination}. 
Hence, for the model in Eq.~\eqref{eq_MFML_examples}, assuming to have $N_{\rm train}^{(5)}=32$ training samples for fidelity $5$ leads to $N_{\rm train}^{(4)}=64$ training samples for fidelity $4$ and 
$N_{\rm train}^{(3)}=128$ training samples on fidelity $3$. 
In Ref.~\cite{zasp19a}, a derivation of the here discussed MFML model from the Sparse Grid Combination Technique 
\cite{harbrecht2013combination,benk2012hybrid,hegland2007combination, hegland2016combination,haji2016multi, Reisinger2012,reisinger2013combination}
has been discussed.
The robustness of this method relies on the assumption that there is a distinct and systematic decrease in the 
error from the respective fidelity to the target fidelity, when increasing the level of fidelity. Previous research shows that this applies to quantum chemistry calculations for atomization energies, first excited state energies, band gaps, and other properties \cite{rasmussen2006gaussian, RupCM, Pilania2017, von2018quantum, ramakrishnan2017machine}.

\subsection{Model Evaluation}
To assess the performance of the constructed ML models, we have to calculate learning curves, which indicate the prediction error for an increasing number of training samples. In all reported results, the learning curves are averaged over a 10-run random shuffling of the training set. For MFML, particular care needs to be taken in the random shuffling to ensure the nestedness of the training samples. 
To this end, first, the training samples for conformations present in the training set $\mathcal{T}^{(F)}$ are shuffled. Then, for each lower fidelity, further samples are consecutively added by a random subset selection process.

Throughout this investigation, all prediction errors have been reported on a holdout set $\mathcal{V}^F:=\{(\boldsymbol{X}_q^{\text{ref}},E^{\text{ref}}_q)\}_{q=1}^{N_{\rm eval}}$, consisting of evaluation representations and the corresponding reference values for the first excited state energies at the target fidelity $F$, i.e., mostly TZVP. No data from the evaluation set is used in any stage of training.
The errors are reported as Mean Absolute Errors (MAEs) which are defined by a discrete $L_1$ norm
\begin{equation}
    MAE = \frac{1}{N_{\rm eval}}\sum_{q=1}^{N_{\rm eval}}\left\lvert P_{\rm ML}\left(\boldsymbol{X}_q^{\text{ref}}\right) - {E}^{\text{ref}}_q\right\rvert~.
    \label{eq_MAE}
\end{equation}
The model $P_{\rm ML}$ can be either identified by the standard KRR model or by the MFML model.

\backmatter
\bmhead{Supplementary information}
Supplementary sections S1-S3, Figs.~S1-S5, and Tables S1-S2. 

\bmhead{Acknowledgements}
The authors acknowledge support by the DFG through the Priority Program SPP 2363 on “Utilization and Development of Machine Learning for Molecular Applications – Molecular Machine Learning” as well as projects KL 1299/24-1 and ZA 1175/3-1.

\bmhead{Declarations}
The authors declare that there is no conflict of interest or any competing interests. 

\bibliography{bibvivinod,ukleine} 

\end{document}